\documentclass[letterpaper,times]{IONconf}
\usepackage{url}

\usepackage[square, comma, sort&compress, numbers]{natbib}

\usepackage[hidelinks]{hyperref}

\usepackage{graphicx}
\usepackage{amsmath,amsthm,amsfonts,amssymb,bm}
\usepackage{float}
\usepackage{subfigure}

\title{Global RTK Positioning in Graphical State Space}

\author{
    Yihong Ge, 
    Sudan Yan,
    Shaolin Lü*,
    Cong Li \\
    \textit{Graph Optimization Inc.}\\
    Beijing, China \\
4050627@qq.com
    }

\begin{document}

\maketitle

\section*{biography}


\biography{YiHong Ge}{graduated from  School of Computer Science, Harbin Institute of Technology, Harbin, China, with a Bachelor's degree in 2019. Since 2019, he worked for Graph Optimization Inc. as a software engineer. His scientific interests include software development for graph optimization in railway surveys and pipeline inspections.}

\biography{SuDan Yan}{graduated from School of Geodesy and Geomatics, Wuhan University, Wuhan, China, with a bachelor's degree in navigation engineering  in 2021. Since 2021, she worked for Graph Optimization Inc. as a navigation engineer. She devotes herself to  the research and development of factor graph optimization navigation algorithm and invariant factor graph optimization framework.}

\biography{ShaoLin Lü*}{received a PHD degree in navigation from Beijing Institute of Technology, Beijing, China, in 2009. From 2009 to 2011, he was with Tsinghua University as a postdoctor. From 2011 to 2018, he was a senior engineer at the NORINCO group. Since 2019, he is the founder of Graph Optimization Inc..}

\biography{Cong Li}{graduated from Department of Precision Instrument, Tsinghua University, Beijing, China, with a Bachelor's degree in 2020. Since 2021, he worked for Graph Optimization Inc. as an FPGA engineer. His scientific interests include FPGA implementation of factor graphs and mechanical structure design.}

\section*{Abstract}
This paper proposes a new  method for RTK  post-processing.
Different from the traditional forward-backward Kalman filter, in our method, the whole system equation is built on a graphical  state space model and solved by factor graph optimization.
The position solution provided by the forward Kalman filter is used as the linearization points of the graphical state space model. Constant variables, such as double-difference ambiguity, will  exist as constants in the graphical state space model, not as time-series variables.  It is shown by experiment results that factor
 graph optimization with a graphical state space model
 is more effective than Kalman filter with a traditional discrete-time state space model  for RTK post-processing problem.

\section{INTRODUCTION}
Global Navigation Satellite System (GNSS) \cite{b1}-\cite{b4} has been widely used in important fields such as surveys, self-driving cars and mobile phones to provide accurate positions. While the single-point positioning technique provides position with meter-level accuracy, Real-Time Kinematic (RTK) positioning technology with centimeter-level accuracy  is usually used in  important scenarios. For some applications, the post-processing RTK algorithm is always a hot issue.

\par In the field of GNSS, state estimation is always an important technique that determines the accuracy of navigation results. As a classical state estimation algorithm, since the emergence of GNSS, Kalman filter \cite{b5}-\cite{b8}  became popular in this field. In the post-processing RTK field, Kalman filter is still dominant \cite{b3}-\cite{b4},   notwithstanding Unscented Kalman filter \cite{b9} and Particle filter \cite{b10} has come forth.
\par Recently, with the improvement of computing power, Factor Graph Optimization (FGO) \cite{b11}-\cite{b12} derived from probabilistic graphical models  \cite{b13}-\cite{b14} has been applied successfully in the field of SLAM (Simultaneous Localization and Mapping), robot control, driverless vehicles and UAV (Unmanned Aerial Vehicle) navigation. There are many examples of Kalman filters that could be replaced by FGO with success  \cite{b12}-\cite{b20}. For this reason, in recent years FGO has become a hit subject and cut-edge technology. Open-source solvers incessantly emerged \cite{b12}, \cite{b19}-\cite{b22}. There are two kinds of graphical models: Bayesian network and Markov field \cite{b14}, and both of them could be transformed into factor graphs. FGO could solve both single-connected graphs and multi-connected graphs, while Kalman filter can only solve a single-connected one. Due to this, unlike Kalman filter, which can only solve the time-series model, factor graph optimization can adopt constant variables in the state space model, which is called the Graphical State Space Model (GSSM) \cite{b23}-\cite{b24}. In Graphical State Space (GSS), for a multi-connected factor graph, the system state at the $k$th epoch can be related to the system state of any epoch. The message passing of the factor graph is bidirectional. Thus, FGO is a natural tool for global data-processing.

\par Kalman filter is essentially a local solver for optimal estimation problems. Generally, if Kalman filter or other forms of optimal estimation algorithm is used, the system will be modeled as a continuous-time system, then it will be discretized as a time-series model which is called Traditional Discrete-Time State Space Model (TDTSSM) here. This model possesses  the Hidden Markov Model property \cite{b14}, which means that the system state at the kth moment is only related to the previous system state at the (k-1)th moment. This property renders the message passing of Kalman filter  merely unidirectional. Therefore, if Kalman filter is used to solve the global data-processing problems \cite{b3}, such as RTK Positioning and  GNSS/INS integrated navigation, Forward-Backward Kalman Filter (FBKF) will be utilized. The data-processing procedure  consists of the steps as follows:\\
(1) Build a continuous-time system model.\\
(2) Discretize the continuous-time system model into a TDTSSM.\\
(3) From the start point of the data, Forward Kalman filter will be used to process the data to the end point of the data in  chronological order. This step is illustrated in Fig. 1.\\
(4) From the end point of the data, Backward Kalman filter will be used to process the data to the start point of the data in  reverse chronological order.\\
This is the standard procedure in global RTK positioning adopted by RTKLIB and other RTK open-source packages \cite{b3}-\cite{b4}. Sometimes, the solution gotten by step $(3)$ and step $(4)$ can be weightedly summed to get a weighted solution.
\begin{figure}[H]
\centering
\includegraphics[width=3in]{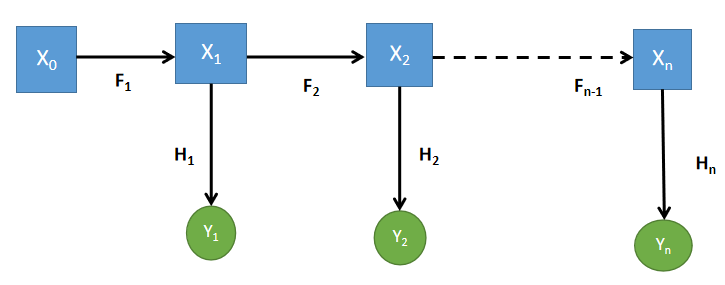}
\caption{State space model representation of Kalman filter.}
\end{figure}
The purpose of this paper is to provide a new solution for global RTK processing which adopts a GSSM solved by FGO. This method has been used in post-processing integrated navigation \cite{b23}. Furthermore, we have utilized this method in other post-processing scenarios, such as railway surveys, pipeline inspections, etc. It has been verified by field test data sampled from a 12km pipeline that GSSM solved by FGO assisted with 2 constrained points can achieve the same accuracy as TDTSSM solved by FBKF assisted with 12 constrained points.

Technically, FGO possesses the ability to optimize both multi-connected graphs and single-connected graphs. It means that FGO could not only optimize TDTSSM which is a single-connected graph, but also optimize a multi-connected graph which describes the evolution of the system state. On the scale of the solvable problems, FGO is a much more powerful optimizer than Kalman filter.
FGO can solve the global data-processing problems in the following steps:\\
(1) Build a continuous-time system model.\\
(2) Discretize the continuous-time system model into a TDTSSM.\\
(3) From the start point of the data, Forward Kalman filter will be used to process the data to the end point of the data in the chronological order. At the same time, store the results of Kalman filter as the linearization points for GSSM.\\
(4) Discretize the continuous-time system model into a GSSM. Build and store the factors which may contain priori factors, measurement model  factors and system prediction model, etc. This step is illustrated in Fig. 2.\\
(5) Use the powerful FGO solver to solve GSSM (a large equation) to get the global optimal solution.

\begin{figure}[H]
\centering
\includegraphics[width=5in]{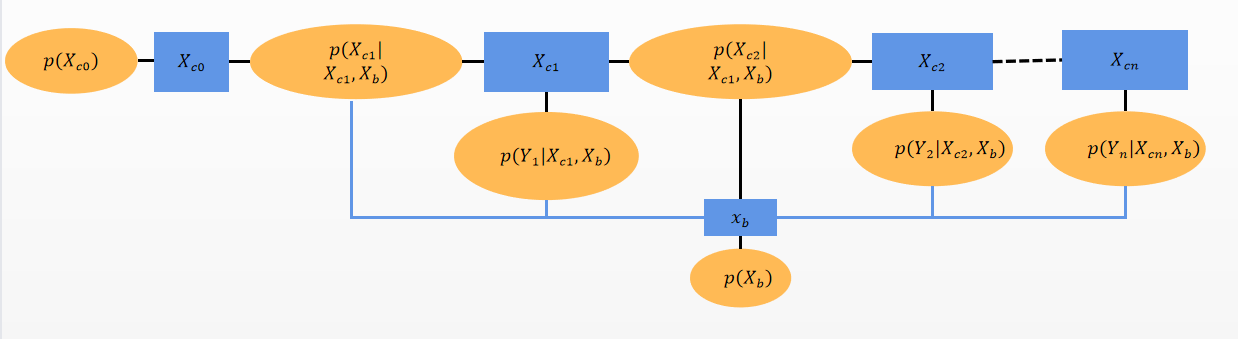}
\caption{Graphical state space model represented in factor graph form for
global optimization.}
\end{figure}

\par For global RTK processing, this method can be used to improve the floating-solution when a fixed-solution can not be gotten. The fix of double-difference ambiguity is an important factor that determines the accuracy of RTK results. However, when the vehicle is in high dynamics or signal blocking is serious, there will be some problems, such as satellite signal out of lock, inefficient observations, multipath effect and cycle slips, etc. Those problems may result in that the double-difference ambiguity could not be fixed. This will be the cause of the decrease in accuracy. By using GSSM solved by FGO, these problems could be alleviated. The fixed solution results of Forward Kalman filter, which consists of position in every epoch and all double-difference ambiguity, are used as linearization points of graph optimization. If at some time, only the float solution is gotten, it will still be used as a linearization point. Different from the TDTSSM solved by  Kalman filter, if a double-difference ambiguity exists in a multi-epoch window, it will propagate as a constant variable, not as a time-series form. The treatment for constant variables is the difference that may improve the accuracy for nonlinear systems \cite{b24}. The clou of the new solution is simple. First, we utilize TDTSSM solved by Forward Kalman filter to get the initialization for GSSM. Secondly,  FGO is used to solve GSSM to get a global optimal RTK solution.
\par This paper is organized as follows. The traditional framework for RTK post-processing, which utilizes TDTSSM solved by FBKF is briefly introduced in section II.
In section III, GSSM is briefly introduced and RTK post-processing solution by GSSM is presented.
In section IV, it is illustrated by experimental results with an open-source implementation that GSSM solved by FGO outperforms TDTSSM solved by FBKF. This paper is concluded with future works in section V.
\section{Traditional Framework For RTK post-processing}
The model in this section is completely the same as the model in RTKLIB, which is very familiar to the community. If you are not familiar with it, check the document of RTKLIB for details \cite{b3}-\cite{b4}. Here it is listed for comparison with GSSM.
\subsection{System Model}
Traditionally, Kalman's discretization framework is adopted in the model of RTK processing. It is assumed that per epoch there is a discrete-time state vector that contains all the variables.
The state model of RTK consists of three parts: position coordinate vector, velocity vector, and the phase bias of carrier wave in different frequencies. The frequencies of satellite data comprise $L_1$, $L_2$ and $L_5$. The state vector can be described as follow:
\begin{gather}
X = [Pos^{T}\,Vel^{T}\,Bias_{1}^{T}\,Bias_{2}^{T}\,Bias_{5}^{T}]
\end{gather}
where $Bias_i=(B_{rb,i}^1,B_{rb,i}^2,...,B_{rb,i}^m)^T$ is single-difference carrier phase biases at $L_i$ frequency.
\par From the viewpoint of a programmer, the dimension of Kalman filter state must be unchanged in the propagation of Kalman filter. In RTKLIB and other open-source packages, the state vector is allocated as a fixed-dimension array no matter
how many satellites are visible. This is one of the limitations of Kalman filter.
\par In RTK positioning, the kinematics model is usually simple and it goes as follows
\begin{gather}
\dot{X}=FX+q=\begin{bmatrix}
0 & I_{3\times 3} & 0& 0&0\\
0 & 0 & 0& 0&0\\
0 & 0 & 0& 0&0\\
0 & 0 & 0& 0&0\\
0 & 0 & 0& 0&0
\end{bmatrix}X+q
\end{gather}
Where $q \sim N(0,Q)$ is the process noise.
\par  The observation variables from the base station and the rover station are used to calculate the double-difference variables. It goes as follows
\begin{gather}
\phi_i = [\phi_{rb,i}^{12}\;\phi_{rb,i}^{13}\;\phi_{rb,i}^{14}\;\cdots\;\phi_{rb,i}^{1m}]\nonumber \\
P_i = [\rho_{rb,i}^{12}\;\rho_{rb,i}^{13}\;\rho_{rb,i}^{14}\;\cdots\;\rho_{rb,i}^{1m}]
\end{gather}
where $\phi$ is double-difference phase-range and  carrier, $P$ is double-difference pseudo-range. $j(2,...,m)$ is the $i$th satellite and superscript $1i(i=2,...,m)$ means the order number of the pair of satellites.  Subscript $rb$ represents rover and base station.
\begin{gather}
\phi_i=
\begin{bmatrix}
\rho_{rb}^{12}+\lambda_i(B_{rb}^1-B_{rb}^2)  \\
\rho_{rb}^{13}+\lambda_i(B_{rb}^1-B_{rb}^3)  \\
\vdots\\
\rho_{rb}^{1m}+\lambda_i(B_{rb}^1-B_{rb}^m)
\end{bmatrix},
P_i=\begin{bmatrix}
\rho_{rb}^{12}  \\
\rho_{rb}^{13}  \\
\vdots\\
\rho_{rb}^{1m}
\end{bmatrix}
\end{gather}
where $\rho_{rb}^{1j}$ represents double-difference pseudo-range of the $1$st satellite and the $j$th satellite between the base and rover station. $\lambda_i$ is the wavelength of  $L_i$ frequency.
As mentioned above, there are triple frequencies, L1, L2 and L5, in the library.
Therefore, the observation vector can be written as
\begin{gather}
Y = (\phi_1^T,\phi_2^T,\phi_5^T,P_1^T,P_2^T,P_5^T)
\end{gather}
The measurement model can be written as
\begin{gather}
Y = h(X)+r
\end{gather}
After linearization, it goes as follows
\begin{gather}
Y = HX+r=\begin{bmatrix}-DE & 0 & \lambda_1D & 0 & 0 \\
			   -DE & 0 & 0 & \lambda_2D & 0 \\
			   -DE & 0 & 0 & 0 & \lambda_5D \\
			   -DE & 0 & 0 & 0 & 0 \\
			   -DE & 0 & 0 & 0 & 0 \\
			   -DE & 0 & 0 & 0 & 0 \\
\end{bmatrix}X+r\\
D =
\begin{bmatrix}1 & -1 & 0 & \cdots & 0\\
			   1 & 0 & -1 & \cdots & 0\\
			   \vdots & \vdots & \vdots & \ddots &\vdots\\
			   1 & 0 & 0 & \cdots & -1\\
\end{bmatrix}\\
E=\begin{bmatrix}e^1_r,e^2_r,...,e^m_r \end{bmatrix}^T
\end{gather}
where  $r\sim N(0,R)$ is the measurement noise. $D$ is a single-difference coefficient matrix. $e^i_r$ is the unit LOS (line‐of‐sight) vector from the receiver to the $i$th satellite.
In fact, both GSSM and TDTSSM are transformed from  the same continuous-time system model described by  Equations $(1)-(9)$. The difference lies in the forms of the discrete-time state space model. In GSSM, not all the variables exist in a time-series form.
\subsection{Forward-Backward Kalman filter}
Generally, Kalman filter is a small-scale framework for the local solution. It can only do the belief propagation in a unique direction. To recursively update the system state, Kalman filter must marginalize out the previous state per update. This makes some information loss from the global viewpoint due to the Hidden Markov Model property.
Before Kalman filter is used, the system (1)-(9) should be  discretized into a TDTSSM as follows
\begin{gather}
X_{k+1}=F_kX_k+q_k, F_k\approx I_{n \times n}+FT\\
Y_{k+1}=H_{k+1}X_{k+1}+r_k
\end{gather}

 The classical Kalman filter can be used to estimate the system  described by Equations $(1)-(11)$.\\
Initialize the  estimation
\begin{gather}
\hat{X}_{0}=E(X_0)=X_0^+ \\
P_0=E[(X_0-\hat{X}_{0})(X_0-\hat{X}_{0})^T]
\end{gather}
Update the prior mean and covariance
\begin{gather}
X_{k+1|k}=F_kX_k\\
P_{k+1|k}=F_kP_kF_k^T+Q_k
\end{gather}
Calculate Kalman gain
\begin{gather}
K_{k+1}=P_{k+1|k}H_{k+1}^T(H_{k+1}P_{k+1|k}H_{k+1}^T+R_k)^{-1}
\end{gather}
Update the posteriori mean and covariance
\begin{gather}
X_{k+1}=X_{k+1|k}+K_{k+1}(Y_{k+1}-H_{k+1}X_{k+1|k})\\
P_{k+1}=(I-K_{k+1}H_{k+1})P_{k+1|k}
\end{gather}

In the post-processing RTK procedure, the data will be processed with TDTSSM solved by  Kalman filter in  chronological order. Then, the floating solution will be gotten. Fixed carrier-phase
ambiguities will be  calculated to improve the accuracy.  However,  the vehicle dynamics or external environment may result in a double-difference ambiguity that could not be fixed. This will be the reason for the decrease in accuracy.
To relieve the unfixed problem, Backward Kalman filter is used. The data will be processed in reverse chronological order with TDTSSM solved by  Kalman filter after the forward process.
 Since Kalman filter can not organize the whole procedure as a large equation, the Forward-Backward method has become a half satisfactory expedient before FGO emerged.

\section{Global RTK post-processing in Graphical State Space}
Generally, TDTSSM follows an implicit assumption: all the state variables are discretized in the same frequency as  time-series forms.
In \cite{b23}-\cite{b24}, it was demonstrated that for some systems this premise may not be followed via GSSM solved by FGO to get better accuracy than TDTSSM solved by  Kalman filter. It should be pointed out that the discretization step from Equations $(2), (7)$ to Equations $(10), (11)$ is not mathematically rigorous or unique. This step is merely a rule of thumb.  There is no doubt that this presupposition is very important for discrete-time Kalman filter and other nonlinear filters. It should be noted that this assumption is usually ignored and supposed to be right by the community.  If you check all the articles or  books on optimal estimation, modern control  and robots, etc., it will be found that this hypothesis is ubiquitous and before \cite{b24} everyone followed it without prior consultation.

The clou of GSSM is to change this step and formulate a  different discrete-time state space model.
\par For the system described by Equations $(1)-(9)$,  there is a different discretization method from that described by Equations $(10)-(11)$. Consider the system state not as a whole, while the constant variables should be discretized as constants, not as time-series forms.  Rewrite the model as follows

\begin{gather}
  \begin{bmatrix}
  \dot{X_c}  \\
  \dot{X_b}
  \end{bmatrix}
=
\begin{bmatrix}
A_c & A_{b} \\
0 & 0
\end{bmatrix}
\begin{bmatrix}
X_c \\
X_b
\end{bmatrix}
+
\begin{bmatrix}
q \\
0
\end{bmatrix} \\
Y=
\begin{bmatrix}
 C_c & C_b
 \end{bmatrix}
 \begin{bmatrix}
  X_c  \\
  X_b
  \end{bmatrix}+r
\end{gather}
Where $X_c$ is the dynamic state vector and $X_b$ is the constant state vector. \\
For the global RTK positioning problem
\begin{gather}
  A_c=0_{3\times 3}, A_b=0\\
  C_c= \begin{bmatrix}
  -DE  \\
  -DE  \\
  -DE  \\
  -DE  \\
  -DE  \\
  -DE \end{bmatrix},C_b=\begin{bmatrix}
  \lambda_1D & 0 & 0 \\
  0& \lambda_2D & 0 \\
  0&  0& \lambda_5D  \\
  0& 0 & 0   \\
    0& 0 & 0   \\
    0& 0 & 0   \end{bmatrix}
\end{gather}

where $X_c = Pos^{T}$ is the state which should be discretized as a time-series form. $X_b = [Bias_{1},Bias_{2},Bias_{5}]^T$ is the constant part that should be
discretized as constants, not as time-series forms. This model is used after the Forward Kalman filter. Due to this, it is not necessary to include the velocity vector in the state vector.
\par When GSSM is built, the whole factor graph consists of the following factors:\\
(1) Priori Factor for the fixed-solution position. After the procedure of Forward Kalman filter, the high accuracy of the fixed-solution position can be used in the factor graph.\\
(2) Double-difference measurement factor. If a double-difference ambiguity exists in multi-epochs, it can be used to improve the low accuracy of floating-solution.\\
The whole factor graph is demonstrated in Fig. 4. The transformation from the model in Fig. 3 to that in Fig. 4 is the key point of GSSM.

\begin{figure} [H]
\begin{minipage}[t]{0.45\linewidth}
\centering
\includegraphics[width=2.5in]{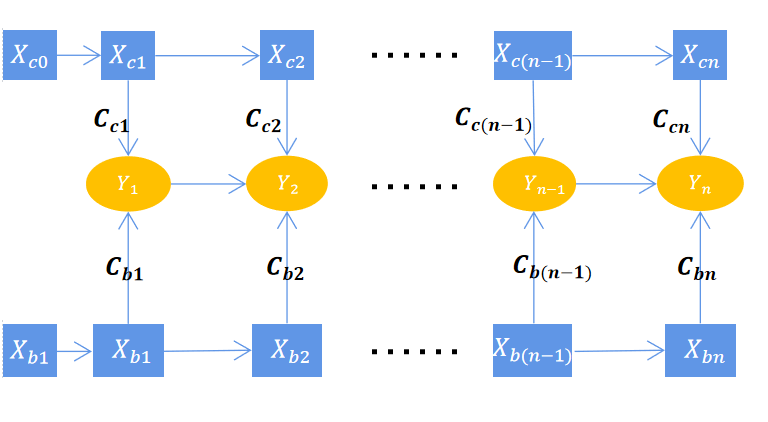}
\caption{Traditional Discrete-time System Model for RTK Post-processing.}
\label{frame}
\end{minipage}%
\begin{minipage}[t]{0.1\linewidth}
\centering
\includegraphics[width=0.5in]{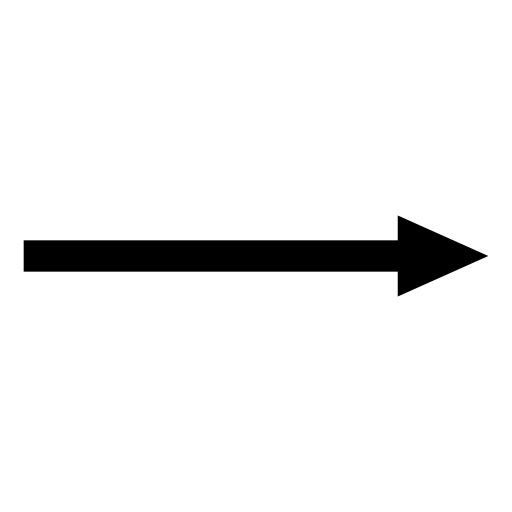}
\end{minipage}
\begin{minipage}[t]{0.45\linewidth}
\centering
\includegraphics[width=2.5in]{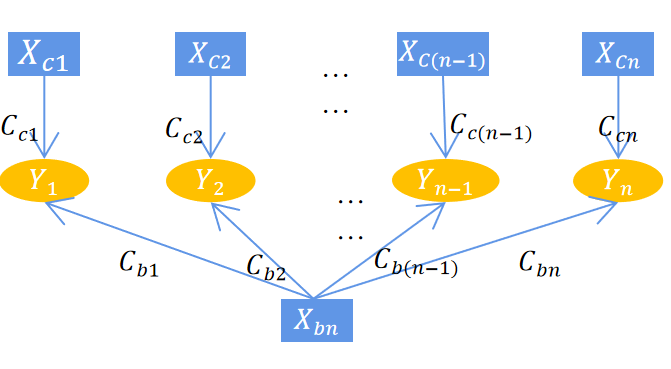}
\caption{Global State Space Model for RTK Post-processing.}
\label{label}
\end{minipage}
\end{figure}
GSSM can be described by a block equation as follows
\begin{gather}
  P_w=diag
\begin{bmatrix}
P_{b}\\
P_1\\
R_1\\
P_2\\
R_2\\
\vdots\\
P_{n}\\
R_{n}\\
\end{bmatrix},
X_w=
\begin{bmatrix}
X_{b} \\
X_{c1}\\
X_{c2}\\
\vdots \\
X_{cn}
\end{bmatrix},
b_w=
\begin{bmatrix}
\hat{X}_{b}^+  \\
\hat{X}_{c1}^+  \\
Y_{1}\\
\hat{X}_{c2}^+  \\
Y_{2}\\
\vdots \\
\hat{X}_{cn}^+  \\
Y_{n}\\
\end{bmatrix},\notag
\\
A_w =
\begin{bmatrix}I & 0 & 0  &\cdots & 0 \\
			   0 & I & 0  &\cdots & 0 \\
			   C_{b1} & C_{c1} & 0 &\cdots & 0 \\
			   0 & 0 & I  &\cdots & 0 \\
         C_{b2}& 0 & C_{c2}  &\cdots & 0\\
			   \vdots & \vdots & \vdots  & \ddots &\vdots\\
			   0 & 0 & 0  & \cdots & I\\
			    C_{bn}& 0 & 0  & \cdots & C_{cn}\\
\end{bmatrix}\notag \\
A_wX_w=b_w
\end{gather}
\par Via FGO,  the solution of Equation (23) can be gotten
\begin{gather}
\hat{X}_w=(A_w^TP_w^{-1}A_w)^{-1}A^T_wP_w^{-1}b_w
\end{gather}
It should be pointed out that this equation can not be directly solved by Kalman filter. The constants must be discretized as time-series forms if Kalman filter is used to solve Equation $(23)$.
The dimension of this equation is unknown before it is formulated for the concrete problem. The variable set solved by Kalman filter can be described as follows
\begin{gather}
(X_{c1},X_{b1})^T,(X_{c2},X_{b2})^T,...,(X_{cn},X_{bn})^T
\end{gather}
Kalman filter can only give the solution of one variable in the above set per update in  chronological order. Meanwhile, GSSM solved by FGO can provide  the solution of the following variable set as a whole per update
\begin{gather}
X_{c1},X_{c2},...,X_{cn},X_b
\end{gather}
Obviously, the solution of the above variable set can not be provided by Kalman filter before $X_b$ is transformed into a time-series form from a constant variable. In Fig.5, the  differences between the two methods are illustrated by a flowing chart.
\begin{figure}[H]
\centering
\includegraphics[width=3.5in]{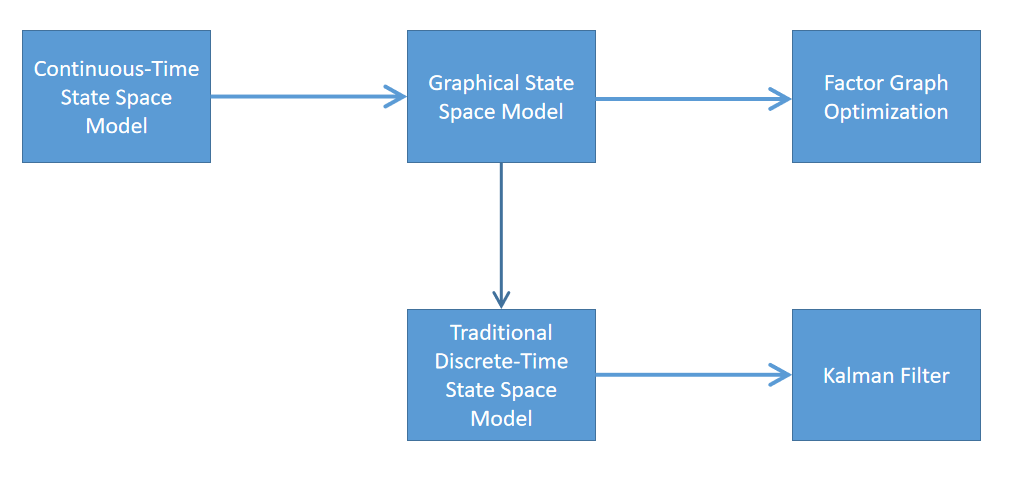}
\caption{The flow chart of the two methods.}
\end{figure}
It should be known that there is an implicit discretization step for the constant variables when TDTSSM is used, this step was not considered before FGO emerged because all nonlinear filters do not possess the ability to solve the equation provided by GSSM. What GSSM does is  construct an equation that can not be solved by Kalman filter. For some kinds of systems, this renders  FGO have the odds to outperform Kalman filter.
\section{Experimental Results}
Field test data is used to compare GSSM solved by FGO with TDTSSM solved by FBKF. Data and the open-source implementation can be found at
GITHUB\footnote{https://github.com/shaolinbit/RTKinGSS}.
It can be concluded that  the accuracy of  floating solution can be improved by the new method.
In RTK positioning, the accuracy of floating-solution in height is notoriously unsatisfactory. In Fig. 6, there is no big difference shown between the two candidate methods in the horizontal direction.
In Fig. 7, it can be seen that the difference lies in the vertical direction when different methods are utilized.

\begin{figure} [H]
\begin{minipage}[t]{0.5\linewidth}
\centering
\includegraphics[width=2.5in]{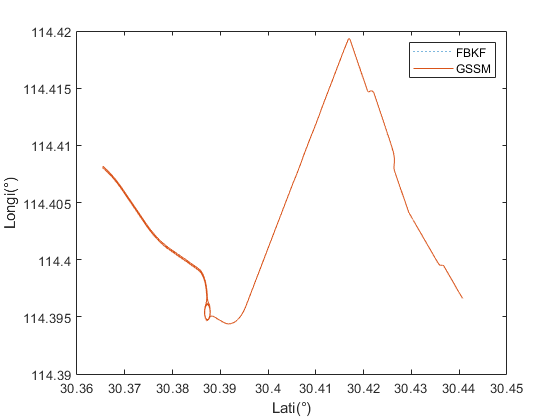}
\caption{Experiment results in the horizontal direction.}
\label{frame}
\end{minipage}%
\begin{minipage}[t]{0.5\linewidth}
\centering
\includegraphics[width=2.5in]{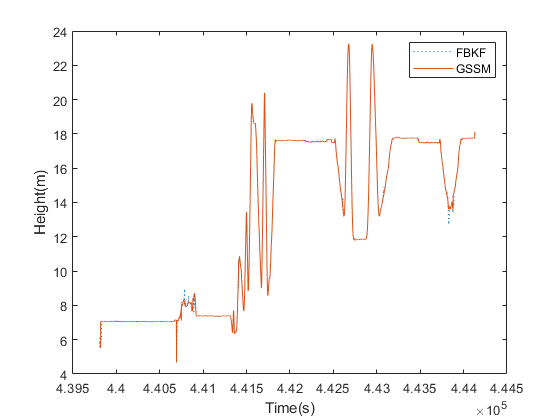}
\caption{Experiment results in the vertical direction.}
\label{label}
\end{minipage}
\end{figure}

From Fig. 8 to Fig. 10, it  can be verified in a challenging  urban canyon when the RTK solution is not fixed, the height solution given by FBKF is unstable. At the same time, the floating solutions given by GSSM are relatively smoother.
\begin{figure} [H]
\subfigure[Comparison GSSM with FBKF.]
{
\begin{minipage}[t]{0.5\linewidth}
\centering
\includegraphics[width=2.5in]{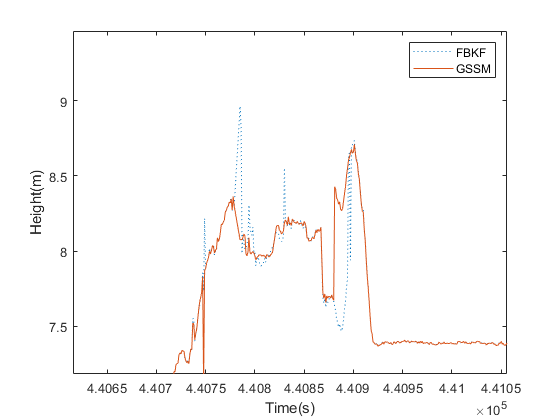}
\label{frame}
\end{minipage}%
}
\subfigure[Solution given by RTKLIB.]
{
\begin{minipage}[t]{0.5\linewidth}
\centering
\includegraphics[width=2.5in]{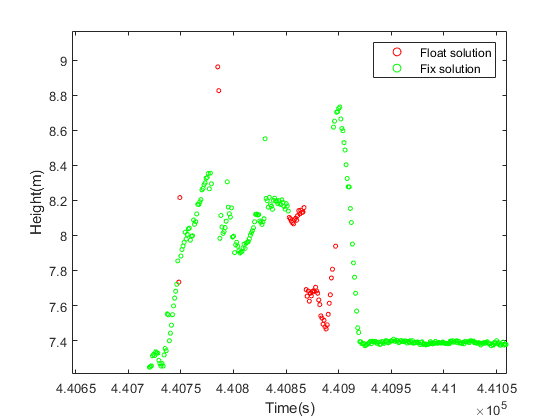}
\label{label}
\end{minipage}
}
\caption{Vertical Positioning results in time interval I}
\end{figure}

\begin{figure} [H]
\subfigure[Comparison GSSM with FBKF.]
{
\begin{minipage}[t]{0.5\linewidth}
\centering
\includegraphics[width=2.5in]{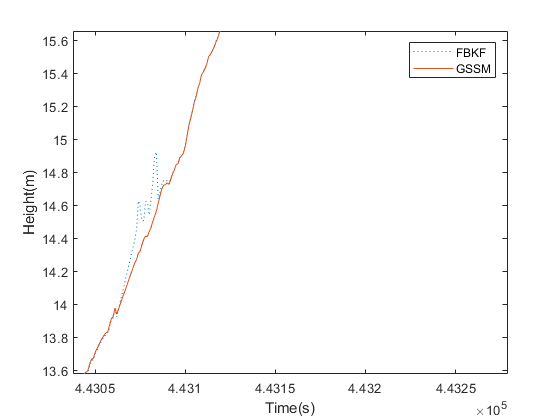}
\label{frame}
\end{minipage}%
}
\subfigure[Solution given by RTKLIB.]
{
\begin{minipage}[t]{0.5\linewidth}
\centering
\includegraphics[width=2.5in]{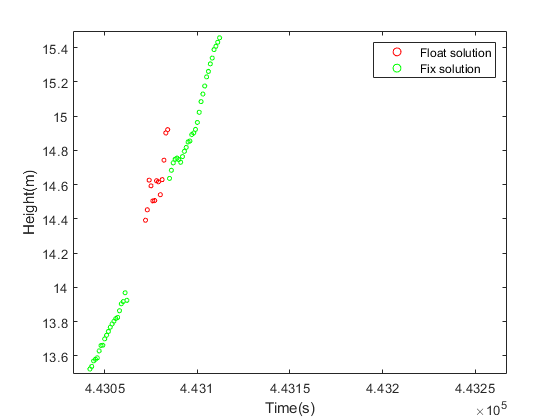}
\label{label}
\end{minipage}
}
\caption{Vertical Positioning results in time interval II.}
\end{figure}

\begin{figure} [H]
\subfigure[Comparison GSSM with FBKF.]
{
\begin{minipage}[t]{0.5\linewidth}
\centering
\includegraphics[width=2.5in]{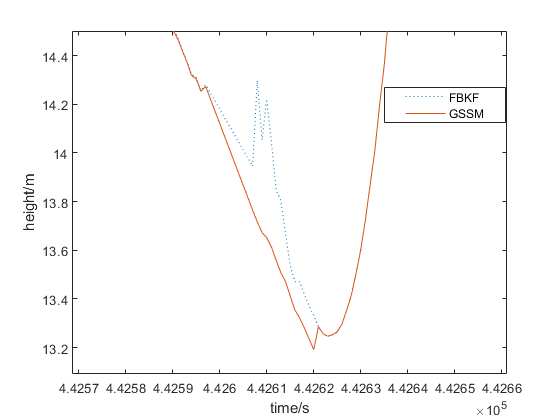}
\label{frame}
\end{minipage}%
}
\subfigure[Solution given by RTKLIB.]
{
\begin{minipage}[t]{0.5\linewidth}
\centering
\includegraphics[width=2.5in]{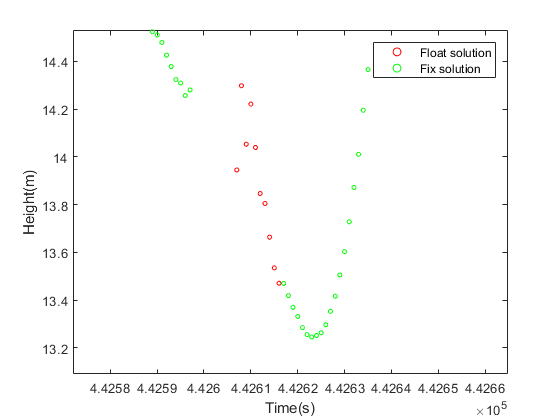}
\label{label}
\end{minipage}
}
\caption{Vertical Positioning results in time interval III.}
\end{figure}

\section{Summary}
A new  method was proposed for RTK  post-processing in this paper.
The whole system equation is built on the graphical  state space model and solved by factor graph optimization, while traditionally forward-backward Kalman filter is used to solve the
equations formulated by the traditional discrete-time state space  model.
Constant variables, such as the double-difference ambiguity, will  exist as constants in the graphical state space model, not as time-series variables.
 It was shown by experimental results that the
proposed method can effectively alleviate the effects
of GNSS outlier measurements, deriving improved accuracy
of floating-solution.  Real-time experiment results and theory analysis will be explored in future works.
\section*{Acknowledgment}
The idea of the Graphical State Space Model for RTK was proposed by Shaolin L{\"u} and he wrote Section III.
The algorithm was firstly implemented by Sudan Yan.

\end{document}